# Do top conferences contain well cited papers or junk?

James Davis

University of California, Santa Cruz

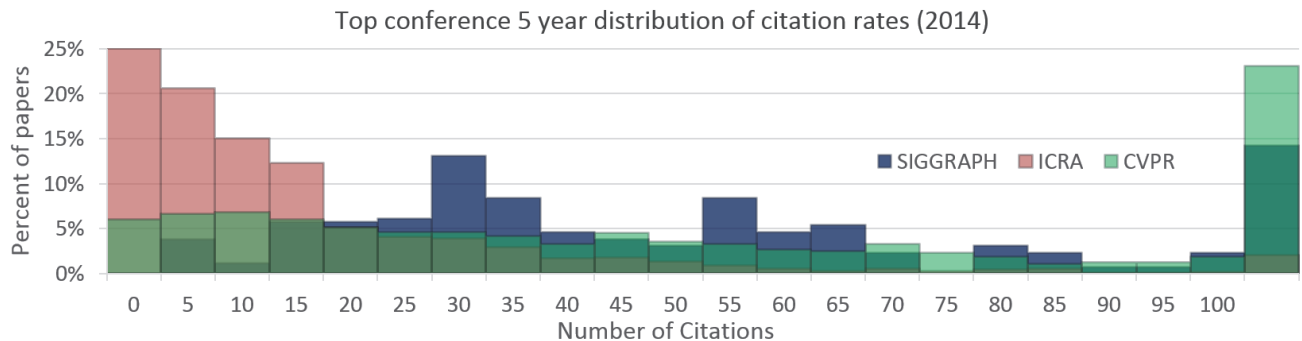

Figure 1: Citations to papers in "top conferences" are clearly drawn from different distributions, as seen in these histograms. ICRA has 25% of papers with less than 5 citations after 5 years, while SIGGRAPH has none. CVPR contains 22% of papers with more than 100 citations after 5 years, a higher fraction than the other two venues.

## Abstract

*In order to answer questions about top conference publication patterns, citation data is collected and analyzed for several computer science conferences, with focus on computer vision and graphics. Both top and second tier conferences are included, and sampling occurred for two different 5 year periods. Example questions include: Do top conferences contain well cited papers or junk? (Yes) Are top conferences similarly cited? (No) Are second tier conferences as good as first tier conferences? (Sometimes) Has something been changing at CVPR? (Yes)*

## 1. Introduction

CVPR has the highest h-index in computer science according to many lists, arguably making it the top conference in CS [30][31][32]. But are all the papers well cited?

H-index measures the number of papers in a conference with high citation [13][14]. This measure implicitly assumes citations as the raw metric for quality. In contrast, CSRankings.org explicitly chooses "top conferences" in each area based on general opinion, and weights publication in any of them equally [3]. Are these top conferences actually equal across areas, or do some contain papers with radically more citations than others?

Certainly researchers do not agree. Any researcher who has been around a few years has likely had a conversation similar to the following:

*Joe Vision*:    CVPR is the best conference!
*Jill Graphics*:    Nah.. CVPR contains a lot of junk!
*Jack Graphics*:    Yo.. SIGGRAPH papers are better than CVPR!
*Joe Vision*:    No way! The top CVPR papers are better than SIGGRAPH!
*Jane Vision*:    Yeah.. CVPR Orals have 4% acceptance! SIGGRAPH is only 20%
*Jill Graphics*:    But …. !

Each individual researcher will have their own anecdotes, and these are passed around in a real world belief propagation network [20][28], which does not seem to converge.

Conference organizing committees also do not agree on their desires for what makes a good conference. Consider these two paraphrased statements made during conference committee planning meetings.

*Joe CVPR*:    *"Accept all good papers, it is ok to accept some lower quality papers if necessary to achieve the goal."*

*Jill SIGGRAPH*:    *"Every paper accepted should be good, it is acceptable to reject some good papers if necessary to achieve the goal."*

These statements were made by luminaries of their respective fields, in the context of a committee discussion of how best to insure the future of their research community. There was no verbal disagreement during the discussion, so one supposes the general sentiments are shared widely within each community, despite the fact they are clearly contradictory.

One is led to wonder about both general questions such as: Are "top conferences" actually equal? and Do citation patterns match the stated goals of communities? As well as more conference specific questions, such as: Are CVPR Orals more impactful than CVPR Posters? Are CVPR Orals more impactful than SIGGRAPH papers? Does CVPR actually contain "junk"?

While the discussions and after hours debates are always good natured, the underlying questions are important. Hiring and tenure letters often reference csrankings.org, number of "top publications", and h-index. What is missing is some hard data to inform the discussion, and the decisions of researchers seeking to submit their work.

In order to answer these questions, we gathered citation rates after 5 years. The data was collected for every paper published in several different conferences, repeating for two different time intervals. Both "top" and "second-tier" conferences were included in the sample.

This paper contributes an analysis of citation rates, comparing top computer vision and graphics venues.

## 2. Related Work

Journal Impact Factor measures the average number of citations for all papers in a venue, and was originally conceived for making journal subscription decisions [8]. It is now used as a shorthand for 'journal prestige' [9], [23]. It is commonly argued that journal level metrics should not be used to judge the worth of individual papers [24], but the practice continues. Under this metric a journal/conference is incentivized to have papers with very high citations and to avoid papers with low citations.

H-index measures the number of papers, N, which have at least N citations, and was originally conceived to measure individual scholars [14]. Hirsch's original paper notes that Nobel Prize winners in physics have a median h-index around 35. Clearly, there is a scaling factor which depends on the author's field of work and the sources used to calculate the metric [1][15]. H-index can also be applied to conference/journal publication lists [4][13]. Under this metric, a high number of papers is rewarded, with no special incentive for papers with very high citations, nor penalty for excess papers with low citation.

There are a wide variety of alternate metrics which have been proposed, often to correct some deficiency in the widely used impact factor and h-index [6][27][29].

Rather than rank journals by a single value, there have been proposals to look at the complete distribution of citations [18][22]. Our work follows this model since it provides a more complete understanding, and allows us to explicitly comment on the frequency of low and high citation papers.

A great deal of work has investigated the important role of conferences in computer science publication, and some have concluded conferences are more impactful than journal publication [11][16][25][26]. In the specific domain of computer vision, Eckmann et al. find a correlation between top-3 conference citation rate and eventual journal publication [5].

This paper explores citation *distributions,* specifically in computer vision conferences.

## 3. Method

Citation statistics were gathered for all papers in multiple conferences, over three time periods. Data was initially collected in 2013 for papers published in 2008. This 5 year citation rate was later augmented by collecting citation rates for the same papers again in 2019, to see if patterns changed significantly with a longer citation period.

Conferences were chosen to cover a few different sub-fields of computer science, and include both top conferences, as well as second tier conferences. The conferences chosen were a convenience sample [10] the authors were familiar with, having heard many informal discussions about the relative (de)merits of these during the last 20+ years of publishing and conference committees.

A second set of data was collected in 2019 for papers published in 2014. Again looking at 5 year citation rates, but changing the list of conferences based on observations and questions raised by the first set of data.

Citation rates for individual papers are those reported by google scholar. The data should be considered noisy since it's well known that authors can game the numbers and have an incentive to do so [2][17][19]. Nevertheless we believe the aggregate findings are valid.

There is no unanimous agreement about what constitutes a "top conference", and any list will have deficiencies, so in this paper we arbitrarily use the list provided by csrankings.org.

All the data and an interactive plot is included in supplemental materials, should the reader want to change plot axes or make comparisons not included here.

## 4. Findings

**Are top conferences in computer science equal?**

Citations to papers in "top conferences" are *clearly drawn from different distributions*, as seen in the histograms of Figure 1. ICRA has 25% of papers with less than 5 citations after 5 years, while CVPR has 22% with more than 100 citations. Histograms are intuitive for most people to interpret rapidly, however since the distributions overlap it is difficult to compare many conferences together. Thus for the remainder of this paper we show data as cumulative distributions, the fraction of papers with at least N citations, as seen in Figure 2. These two figures have exactly the same data, presented in different formats.

**Has CVPR changed recently?**

*CVPR papers are now cited much more than they were in the past*. Figure 3 compares Poster and Oral papers in 2008 and 2014. In 2008, Poster papers had a median of 19 citations after 5 years, but by 2014 the median 5 year citation rate had increased to 43. The increase in citations to Oral papers is not as strong, but still clear.

The number of papers receiving high citations has also doubled among both posters and orals. In 2014 23% of all CVPR papers had more than 100 citations after 5 years.

The changes are even more extreme among the citation rich. In 2008 none of the seven conferences surveyed had any papers with more than 1000 citations after 5 years. In 2014 CVPR and ECCV each had 7 papers with more than 1000 citations. The top paper at CVPR had 10,000 citations, more than being cited by *every* CVPR, ICCV, ECCV, ACCV, and BMVC paper that has come after!

**Are CVPR Orals better than Posters?**

*In 2008, the median CVPR Oral paper had more than twice as many citations as the median CVPR Poster*, as seen in Figure 3. Looking at only papers with more than 100 citations, we see that 22% of all CVPR Orals reach this bar, while only a smaller fraction of CVPR Posters achieve this level of fame.

*However it would be wrong to think of these as separable categories*, with all Orals better than all Posters. First, citation histograms overlap with both categories containing papers with few citations as well as papers with many citations. Second, Posters from 2014 have almost the same 5 year citation distribution that Orals had in 2008.

**Do top conferences contain junk papers?**

*CVPR does contain papers with low citations*. More than 10% of papers receive fewer than 10 citations, and 2% have 0 citations after 5 years, as shown in Figure 2. However top vision conferences accept thousands of papers, allowing many researchers to participate. This appears to match the community's stated goal of "accepting all good papers, allowing for some poor papers to achieve this".

ICRA has more weak papers with more than 40% of papers receiving fewer than 10 citations after 5 years.

SIGGRAPH has no papers with 0 citations, and less than 2% have fewer than 10 citations after 5 years. This appears to match the community's stated goal of "accepting only good papers". However this comes at the cost of excluding much research from top conferences, since SIGGRAPH accepts only about 100 papers a year.

All of these top conferences have better citation rates than the general scientific literature, for which it is estimated that between 10% and 50% of papers remain completely uncited after 5 years [12], [21]. As an (imperfect) comparison to a top journal outside computer science, Nature has 12% of papers with fewer than 10 citations after 2 years [18].

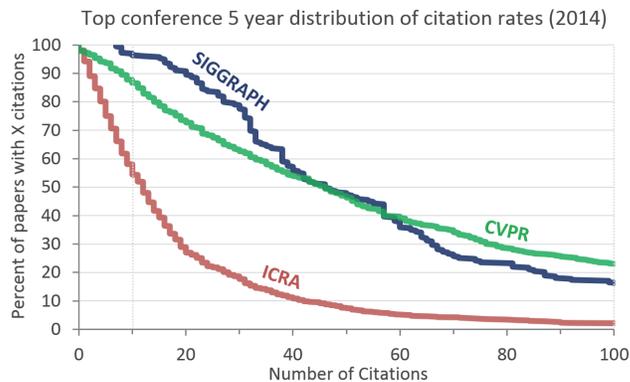

Figure 2: The citation distributions of SIGGRAPH, CVPR, and ICRA are shown as cumulative distributions, the fraction of papers with at least N citations. Note that ICRA has 40% of papers with fewer than 10 citations, CVPR 10%, and SIGGRAPH less than 2%, yet all are "top conferences".

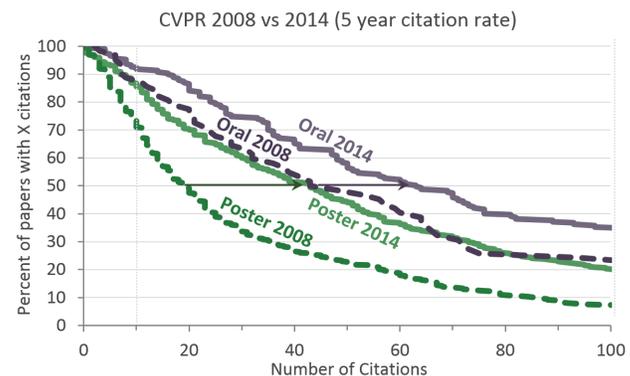

Figure 3: CVPR papers are compared for two time periods. In both cases CVPR Orals receive more citations than do Posters, however low and high citation papers exist in both categories. Interestingly, CVPR citations increased substantially over time. This effect is so great that later CVPR Posters received as many citations as earlier CVPR Orals.

**Are second tier conferences as good as top conferences?**

*Second tier conferences often do have a citation distribution similar to top conferences*.

BMVC is a well-known general computer vision conference, and ICCP is a smaller specialty conference on computational photography. Both are well respected but second tier. As shown in Figure 4, these conferences have citation patterns very similar to ICRA, which is a top conference in robotics.

Eurographics, I3D, and SCA similarly span general and specialty second tier graphics conferences. The median paper at these venues in 2008 received more citations than the median CVPR paper, as shown in Figure 5. If we look at 10 year citation rate, shown in Figure 6, the conclusion is similar. The primary difference is that SCA and Eurographics now include a noticeable fraction of high citation papers, something they were lacking after only 5 years.

**Is SIGGRAPH better than TOG?**

ACM Transactions of Graphics (TOG) and "SIGGRAPH in Asia" are an interesting case study. The graphics community long maintained that SIGGRAPH was better than the best graphics journal, ACM TOG. To rectify this inversion, which was confusing to other scientists who value journals more, SIGGRAPH was relabeled as a special issue of TOG. When ACM created a second premiere graphics conference, "SIGGRAPH in Asia" in 2008, this was also considered 'not as good' by many. However this second SIGGRAPH conference is now considered a top conference, and one wonders if the community bias has remained.

The data shown in Figure 7 supports the belief that *SIGGRAPH papers receive more citations than do papers submitted to SIGGRAPH in Asia or directly to the journal TOG*, despite the fact that these are all formally cited as TOG publications. Interestingly, the overall citation rates for TOG and SIGGRAPH in Asia have not changed since 2008. Instead the citation rates of the original SIGGRAPH have declined somewhat. This is the opposite of what occurred at CVPR over the same time period.

**How many citations does a paper in Y probably have?**

Suppose we define "probably" as the middle 50% of papers, among neither the best nor the worst. Figure 8, shows the number of citations a paper probably has, with conferences sorted by median citations.

## 5. Discussion

This paper explores the relationship between top conferences and citation rates. Implicit in much of the discussion is that better papers have more citations. This is of course far from a strict relationship. There are a great many factors possibly impacting citation rate, including the popularity of the topic, the fame of the authors, and the publication venue. It is not our intent to argue that citations are a *better* indicator of quality than publication in a top conference. Citation is merely a different indicator. Nevertheless it seems reasonable to assert that papers with very few citations are not impacting the research community much, and those with a very high number have had some impact.

The analysis in this paper suggests that variations across papers within any conference are so large as to make evaluation based solely on publication venue meaningless. It is also wrong to compare publication at a top conference in subfield A to a top conference in subfield B. The differences are simply too great. This is similar to the known problem comparing citation counts and h-index across fields.

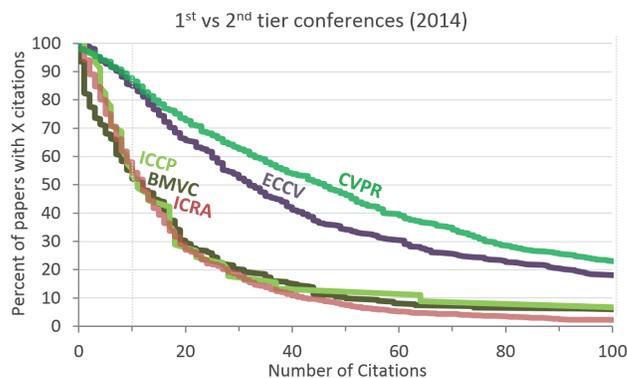

Figure 4: Top conferences (CVPR, ECCV, ICRA) are compared to two second tier conferences (BMVC, ICCP). Note that the citation distribution for these second tier conferences matches the distribution for one of the top conferences very closely.

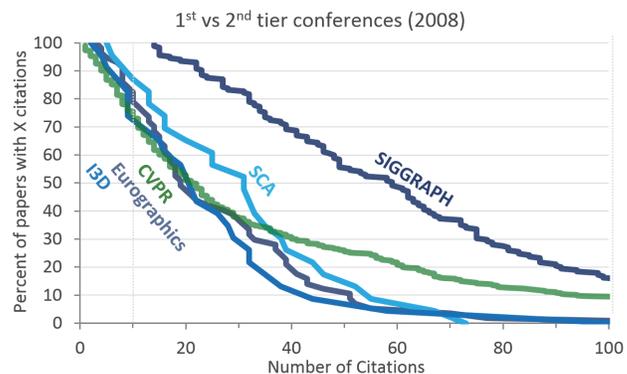

Figure 5: Top conferences (CVPR, SIGGRAPH) are compared to three second tier conferences (SCA, Eurographics, I3D). Note that the median citation rate after five years for these second tier conferences is slightly above the median citation rate for CVPR.

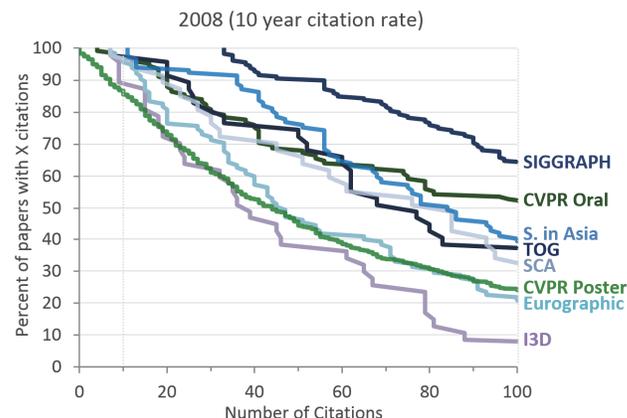

Figure 6: Top conferences (CVPR, SIGGRAPH, SIGGRAPH in Asia) are compared to three second tier conferences (SCA, Eurographics, I3D) for a longer citation period. Note that the ten year citation distributions for SCA and Eurographics fall between the distributions for CVPR Orals and Posters, a top conference.

One might wonder if acceptance rates at conferences is a good indicator, but the analysis here suggest it is badly correlated with citation. CVPR Orals in 2008 with acceptance of 4% had roughly the same citation distribution as CVPR Posters in 2014 with acceptance of 30%. This matches prior studies that conclude acceptance rate is not an indicator of eventual citation rate [7].

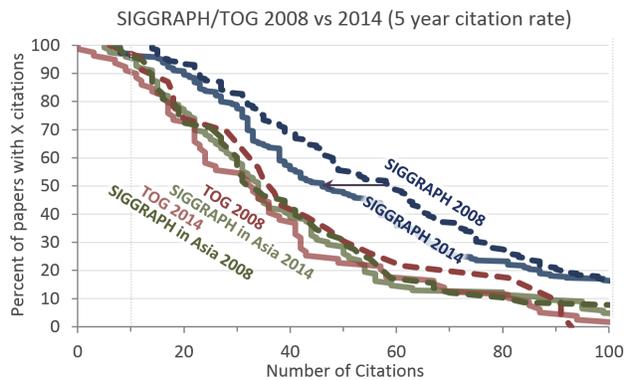

Figure 7: The citation rates of the conference SIGGRAPH are compared with direct submissions to the journal ACM TOG. Note that the conference submissions had a higher citation distribution, even though all are formally cited as published in TOG. However, this difference declined between 2008 and 2014.

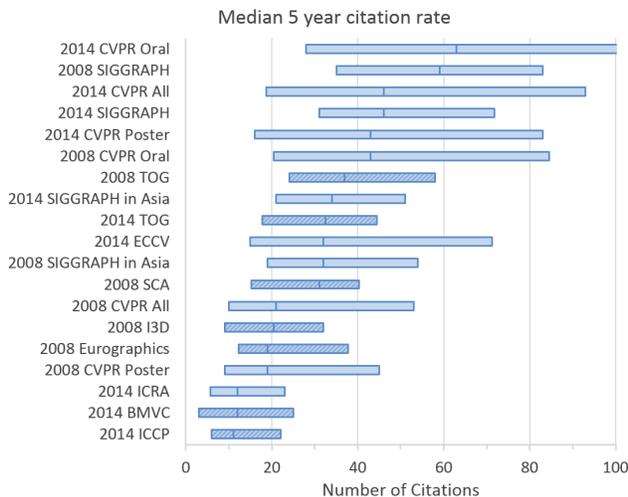

Figure 8: The number of citations a medium quality paper has if published in conference Y. The visualization is sorted by median citations, and shows bars extending to the 25% and 75% most cited papers. Top conferences have solid bars, and second tier conferences have hashed bars.

Perhaps the most surprising finding to the authors is the huge growth in citations that has occurred at CVPR, and the phenomena of some papers having ridiculously high citation rates. This was not mirrored in graphics, the other subfield investigated. Nor was it mirrored when looking at all studied conferences holistically, as shown in Figure 9. Given the reliance of many evaluators on automated citation metrics produced by Google Scholar, this "grade inflation" is likely having a positive impact on the careers of many computer vision researchers. Has something fundamental changed in the field to justify this inflation? Or is this merely an artifact of changes in paper citing practices?

Seglen analyzes the direction of causality and concludes that venue is not the cause of high citation, but rather highly cited papers produce good venues [24]. However this does not explain the growth of citations at CVPR. It seems easier to believe that the conference has changed, than to believe that all of the authors have started writing more impactful papers.

If there is any real conclusion to this study, it is that statements like "She had 10 papers this year in top conferences!" and "He has 10,000 citations!" are at best incomplete, and at worst irresponsible. Certainly *you* knew that already. You're the kind of person who reads research in order to evaluate it. But I bet you can think of at least one administrator, granting agency, or hiring committee which resorts to simple metrics to evaluate work. This paper seeks to either help you satisfy the simple numeric requirements more efficiently, or argue against them more effectively, according to your preference.

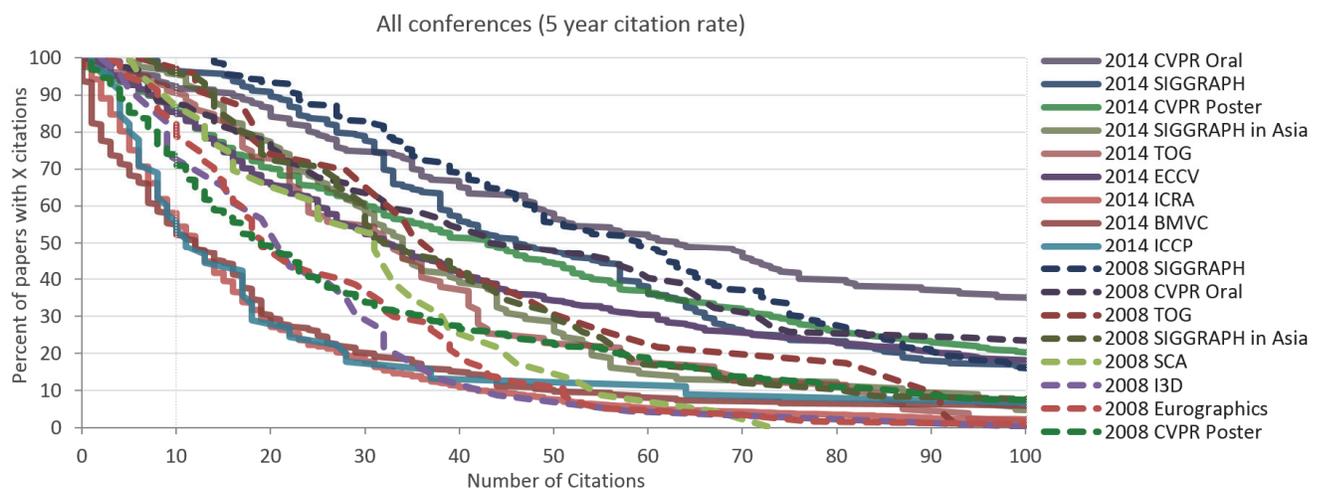

Figure 9: All conferences shown together with 5 year citation rates. All conferences have a wide range of papers, with citations from low to high. There is no obvious overall pattern of more citations during the later time period.